\documentclass[trackchanges]{aastex7}

\begin{document}

\title{Understanding the complex morphology of a CME: \\multi-view analysis and numerical modeling}

\author[orcid=0009-0009-9799-979X]{Cecilia Mac Cormack}
\affiliation{Heliospheric Physics Laboratory, Heliophysics Science Division, NASA Goddard Space Flight Center, 8800 Greenbelt Rd., Greenbelt, MD 20770, USA}
\affiliation{The Catholic University of America, Washington, DC 20064, USA}
\email[show]{cecilia.maccormack@nasa.gov}  

\author[orcid=0000-0001-5400-2800]{Abril Sahade} 
\affiliation{Heliospheric Physics Laboratory, Heliophysics Science Division, NASA Goddard Space Flight Center, 8800 Greenbelt Rd., Greenbelt, MD 20770, USA}
\email[show]{abril.sahade@nasa.gov}

\author[orcid=0000-0002-8164-5948]{Angelos Vourlidas} 
\affiliation{The Johns Hopkins University Applied Physics Laboratory, Laurel MD 20723, USA.}
\email{Angelos.Vourlidas@jhuapl.edu}

\author[orcid=0000-0003-1377-6353]{Phillip Hess}
\affiliation{U.S. Naval Research Laboratory, Washington, D.C., USA}
\email{phillip.n.hess2.civ@us.navy.mil}

\author[orcid=0000-0002-3253-4205]{Robin Colaninno}
\affiliation{U.S. Naval Research Laboratory, Washington, D.C., USA}
\email{robin.c.colaninno.civ@us.navy.mil}

\author[orcid=0000-0003-0565-4890]{Teresa Nieves-Chinchilla}
\affiliation{Heliospheric Physics Laboratory, Heliophysics Science Division, NASA Goddard Space Flight Center, 8800 Greenbelt Rd., Greenbelt, MD 20770, USA}
\email{teresa.nieves@nasa.gov}

\begin{abstract}
Although all coronal mass ejections (CMEs) that propagate into the heliosphere should contain a magnetic flux rope (MFR) component, the majority do not exhibit the expected white-light MFR morphology of a leading edge plus cavity. This different appearance could be the result of distortion of the internal magnetic structure, merging with other structures, or simply projection effects. These factors complicate the interpretation of CMEs. 
This complexity is exemplified by a CME observed on 28 March 2022. The event originated from a single eruption, evolving as a textbook CME in the low corona but appearing as a complex two-MFR structure in white-light observations. Why?
To answer this question, we performed a multi-view data and modeling analysis to describe the CME coronal evolution. {The thermodynamic MHD model, CORHEL-CME, helps reveal the magnetic configuration of this CME and also reveals that the ambient field plays a crucial role in shaping the complex structure of the CME during early evolution.} Our research underscores the importance of integrating multiview observations with physics-based models to gain a deeper insight into the development of {complex} CMEs.

\end{abstract}




%-------------------------------------------------------------------

\section{Introduction} \label{sec:intro}

Coronal mass ejections (CMEs) are the main drivers of space weather disturbances. They originate from the eruption of a stressed magnetic system and evolve throughout the solar corona into the interplanetary medium \citep[e.g.,][]{patsourakos_2020}. Understanding their evolution and three-dimensional (3D) magnetic structure is crucial to predict and {mitigate the} impacts that can damage space instrumentation and communications. Although the magnetic field is the most impactful aspect of the CME in space weather, remote sensing can only image the plasma density structures of the CME. For most of the CME evolution, the magnetic structure has to be inferred from the density. Even when direct in situ measurements of the magnetic field are available, this data only samples a small part of the global structure.  

The basic magnetic structure of CMEs is widely accepted to be a magnetic flux rope (MFR) \citep[e.g.,][]{vourlidas_2014}. In white-light observations, this MFR structure is typically simplified as a three-part structure consisting of a leading edge, cavity, and a bright core {\citep{illing_1985,hutton_2015}}. The common ``light-bulb'' depiction of the three-part structure assumes a side view of the MFR that is only one possible orientation of the MFR 3D structure. Projection effects can change the appearance of the MFR without violating the basic assumption of the model. 

Although many examples of three-part CMEs have been reported and analyzed in the literature, especially between 2 and 20$\,$R$_{\odot}$, the majority of CMEs deviate from this simplified morphology. These deviations can be attributed to various factors. CME identification can be complicated by obscuration or interaction with surrounding structures. Interactions between CMEs and the strong ambient magnetic field in the corona can cause significant deformations, distortions, and deflections \citep[e.g.,][]{kay_2017,sahade_2023}, leading to misinterpretation of the final CME structure. These complexities make it challenging to prescribe a physically realistic magnetic structure to white-light CME observations.

To achieve a consistent 3D description of the evolution of the magnetic structure of CMEs observed in EUV and white-light, several reconstruction tools and models have been developed. Among the simplest, the Graduate Cylindrical Shell \citep[GCS,][]{thernisien_2006, thernisien_2011} assumes a flux-rope shape to model the position and orientation of the CME. This forward modeling tool is helpful for distinguishing between the complexity of projection effects from different {points of view (POVs)} and the real complexity of the magnetic system, despite only containing the geometry of a simplified CME. Because the model is idealized, it cannot account for distortions and complexity within an event, and including such features would require the use of many more free parameters within the geometry. A common solution for more complex events is to use more than one GCS for a single eruption \citep[see e.g., ][]{thernisien_2009, rodriguez-garcia_2022} to model individual parts of the event. While sometimes complex events are the result from multiple eruptions \citep{colaninno_2015}, this approach can lead to a misunderstanding of the internal MFR structure and affect space weather predictions.

Magnetohydrodynamic (MHD) models are able to simulate the interaction between CMEs and the magnetic environment in which they are evolving \citep{cargill_2002, riley_2003, verbeke_2019} to obtain a more realistic description of an eruptive CME at the {expense of computing time}. In particular, thermodynamic MHD models incorporate information about the energy transport processes around an explosive event. These results can {more closely match} EUV/X-rays and white-light observations \citep[e.g.,][]{lionello_2009, downs_2013, torok_2018} to help clarify the relationship between the observations and the underlying magnetic field.

%Because of the importance of the magnetic field dynamics in governing CME propagation and evolution, and the difficulty in observing this field directly, the necessity of combining high-resolution observations from multiple {POVs} with physics-based modeling is clear. This synergy enables a more comprehensive understanding of the internal magnetic configuration of CMEs and their evolution within the inner heliosphere. 

Due to the critical role magnetic field dynamics play in governing the CME propagation and evolution, and the lack of direct observations of it, the integration of high-resolution observations from multiple POVs with physics-based modeling is essential.

New solar missions, offering high-resolution observations from new POVs, {have enabled a} more comprehensive coverage of CME evolution.
The Solar Orbiter mission \citep[SolO,][]{muller_2020}, includes the Solar Orbiter Heliospheric Imager \citep[SoloHI;][]{howard_2020}, which has a resolution capable of detecting fine structures in CMEs. SolO is an inner heliospheric mission in an elliptical orbit. It provides a unique vantage point that can significantly improve the understanding of the global CME structure when combined with observations from imagers on spacecraft at 1$\,$au, such as the Solar and Heliospheric Observatory \citep[SOHO]{domingo_1995} and the Solar TErrestrial RElations Observatory \citep[STEREO]{kaiser_2008}, and numerical models.

The improved resolution capabilities of SoloHI is demonstrated in \citet{hess_2023}. In this paper, the authors compared CMEs observed simultaneously with the inner heliospheric imager (HI1) of the Sun-Earth Connection Coronal and Heliospheric Investigation \citep[SECCHI]{howard_2008} suite on board the STEREO-A and SoloHI. As a particular example, the authors compared the March 28 2022 (hereafter March 28) event, which exhibited a complex configuration in all imagers. While HI1 sees a distorted CME, SoloHI resolves two bright fronts that can be attributed to either a highly distorted CME and shock or to two unrelated structures propagating along the same line of sight. Moreover, the coronagraphic observations from 1$\,$au suggest the appearance of two CMEs or, at least, two MFR orientations. 
The early eruption and source region of the March 28 event were also well-observed due to its central location on the solar disk from Earth's POV. These observations are well-complemented by the ones provided by STEREO and SolO, located at $-33.3^{\circ}$ and $83.9^{\circ}$ from the Sun-Earth line, respectively. 
With this complimentary multi-viewpoint coverage, the source region dynamics indicate a single eruption.

{In this study, we undertake a deep analysis of the March 28 CME to determine its magnetic nature by leveraging the complementary coverage provided by SolO, STEREO-A and SOHO and modeling the CME evolution with the CORHEL-CME thermodynamic MHD model \citep{linker_2024}. }From the simulations, we provide an explanation for why this single eruption evolved into a CME with a complex morphology in the white-light observations. This study {presents} a more comprehensive understanding of the internal structures of this CME and its interpretation{ in observations, and provides }a potential framework applicable to more events.

In Section \ref{sec:data} we describe the white-light observations. In Section \ref{sec:source+CORHEL} we analyze the source region and filament eruption and use this information to initiate the MHD simulation. The results and the comparison of the model to the observations are discussed in Section \ref{sec:Model}. We conclude in Section \ref{sec:conc}.

\section{Observations} \label{sec:data}

\begin{figure*}[ht!]
\epsscale{1.15}
\begin{interactive}{animation}{20220328_CME.mp4}
\plotone{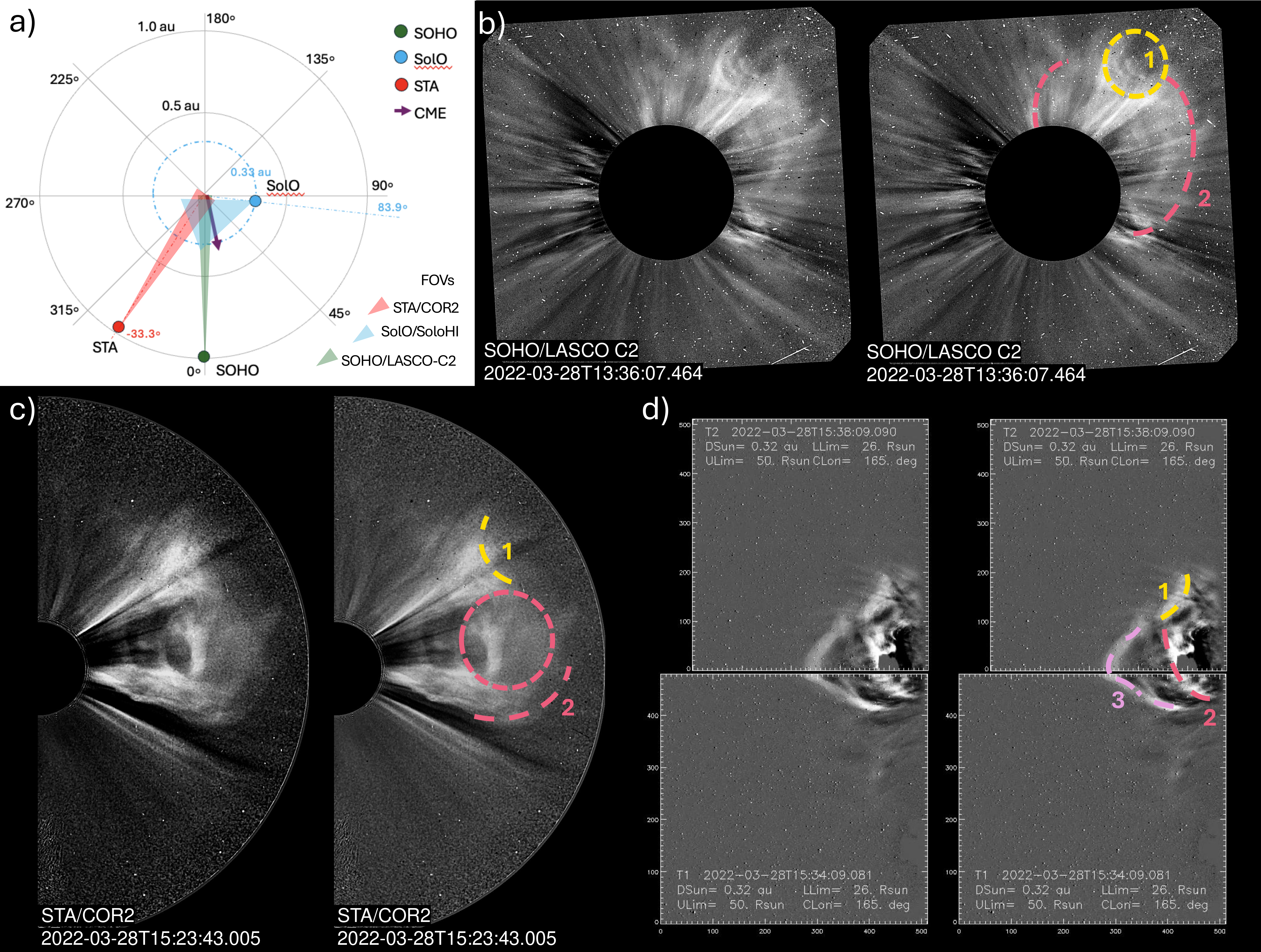}%{RS_obs3.pdf}%{Multi_RS.png}
\end{interactive}
\caption{a) Spacecraft configuration on March 28, 2022. b) LASCO C2 base difference images for the March 28 event at 13:36:07, numbered dashed curves denote features mentioned in the main text. c) Same of panel b but for COR2 at 15:23:43. d)  Same of panel b but for the SoloHI inner FOV processed with running difference. {The CME evolution movies for the three shown FOVs are included as an animated version for this figure. The linked movie starts with only the CME evolution in the LASCO-C2 FOV, the movie for the COR2 FOV begins around 00:01 and the CME evolution in the SoloHI FOV begins around 00:03.} \label{fig:LC2_0328}}
\end{figure*}

The March 28 CME is well observed in white light imaging from three POVs. The source region of the CME is at {14N/4W relative to Sun-Earth line} and, based on the observations, is Earth-directed. Figure \ref{fig:LC2_0328}a shows the location of STEREO-A (STA), SolO and SOHO at on 28 March 2022 with respect to the CME propagation direction. STEREO and SolO are located at $-33.3^{\circ}$ and $83.9^{\circ}$ from the Sun-Earth line, respectively and proved unique side views of the CME.   

The CME was observed by SoloHI, the Large Angle and Spectrometric Coronagraph \citep[LASCO][]{brueckner_1995} on board SOHO (hereafter LASCO), and the COR2 coronagraph on board the STEREO-A spacecraft (hereafter COR2). The event is listed in the multi-viewpoint \href{https://science.gsfc.nasa.gov/lassos/ICME_catalogs/solohi-catalog.shtml}{SoloHI catalog}, which includes data provided by other heliospheric missions to describe each event observed by SoloHI \citep{maccormack_2025}.

The CME first appeared in the LASCO/C2 FOV at $\sim$12:00$\,$UT (Figure \ref{fig:LC2_0328}b). As it propagates through the FOV, the CME appears as a partial halo with a ragged front (dashed pink curve 2 in Figure \ref{fig:LC2_0328}b). A U-shaped cavity (dashed yellow curve 1 in Figure \ref{fig:LC2_0328}b) propagates out at a position angle of $\sim300^{\circ}$ and can be tracked throughout the FOV. %can be clearly detected in the SOHO/LASCO C2 FOV and it remained stable throughout the CME's evolution. 
The U-shape morphology is a common MFR white-light signature when observing a CME from the side, directly along its central axis. For this event, the location of this feature relative to the halo front makes it difficult to determine whether it is part of the same CME or is a distinct eruption. If it is the latter case, we would also want to understand the relationship between this smaller structure and the halo, since they propagate at a similar rate. 

The CME appears in COR2 at 12:08$\,$UT. Figure \ref{fig:LC2_0328}c shows the CME at 15:23$\,$UT with a diffuse leading edge (dashed pink curve 2) and a complex core with a faint cavity (dashed pink circle), oriented largely perpendicular to the plane of the sky. These components of the CME are consistent with an Earth-directed MFR-CME as viewed from the STA perspective, $\sim33^{\circ}$ east of the Earth. The U-shape structure from the C2 images is also visible in COR2, as a concave front (dashed yellow curve 1). Again, the COR2 images could be interpreted as signatures of two CMEs, one heading to Earth and another (or a component of the Earth-directed CME) heading to the north. 

Structures 1 and 2 are also detected in the SoloHI observations (Figure \ref{fig:LC2_0328}d). However, a v-shaped front (marked `3') is seen ahead of structure 2. The March 28 CME entered the SoloHI FOV at approximately 13:00$\,$UT. Figure \ref{fig:LC2_0328}d shows the event at 15:34:09~UT in the two inner tiles of the SoloHI FOV using a running difference process to highlight the CME. As the CME evolve through the SoloHI FOV, the highlighted structures remain distinct. {The additional v-shape front (dashed violet curve 3 in Figure \ref{fig:LC2_0328}d) does not fit with a simplified MFR model of the CME structure either.} This poses a new question about the origins of this third front. Between 15:40 and 19:50 UT, there is a data gap in the SoloHI images, as the instrument stopped observing during a planned spacecraft maneuver.

%Front 3 has a v-shaped form (dashed violet curve 3 in Figure \ref{fig:LC2_0328}d), followed by the fronts identified in the other imagers. 

To understand the origins of the CME appearance and its magnetic structure, we need to analyze its low corona evolution and look for hints to the processes that led to any deformation. We simulated this event with a thermodynamic MHD model. We correlated the evolution of the internal magnetic field and its interaction with the environment. We use the CORHEL-CME tool in the CORona-HELiospheric (CORHEL) modeling suite \citep{linker_2024}, which allows the simulation of individual CME events using the photospheric magnetic field as a boundary condition. This tool is available at the Community Coordinated Modeling Center (\href{https://ccmc.gsfc.nasa.gov}{CCMC}) to run-on-request. 

The source region for this event is AR12975 (SHARP region 8088), located near disk center on March 28. The \href{https://solarmonitor.org/full_disk.php?date=20220328&type=shmi_maglc&region=&indexnum=1}{Solar Monitor webpage} lists a large number of flares originating in this AR, including an M4.0 class flare starting at $\sim$10:58$\,$UT and peaking at 11:29$\,$UT. This flare is associated with the filament eruption and the March 28 CME, and there is no reported evidence supporting a double MFR eruption scenario \citep[see e.g.,][]{purkhart_2023,morosan_2024,podladchikova_2024}. {\cite{sahade_2024} investigated} the evolution of the filament and the magnetic environment surrounding the source region of the March 28 CME. We identified a helmet streamer structure overlying the filament channel that may have induced deflection and rotation of the erupting filament. Note that AR 12975 is also surrounded by two coronal holes to the east (negative polarity) and west (positive polarity) sides. 

The next section presents the analysis of the source region and considerations to obtain an optimal reproduction of the eruption with CORHEL-CME.

\section{Analysis}
\label{sec:source+CORHEL}

\begin{figure*}[ht!]
\epsscale{1.00}
\plotone{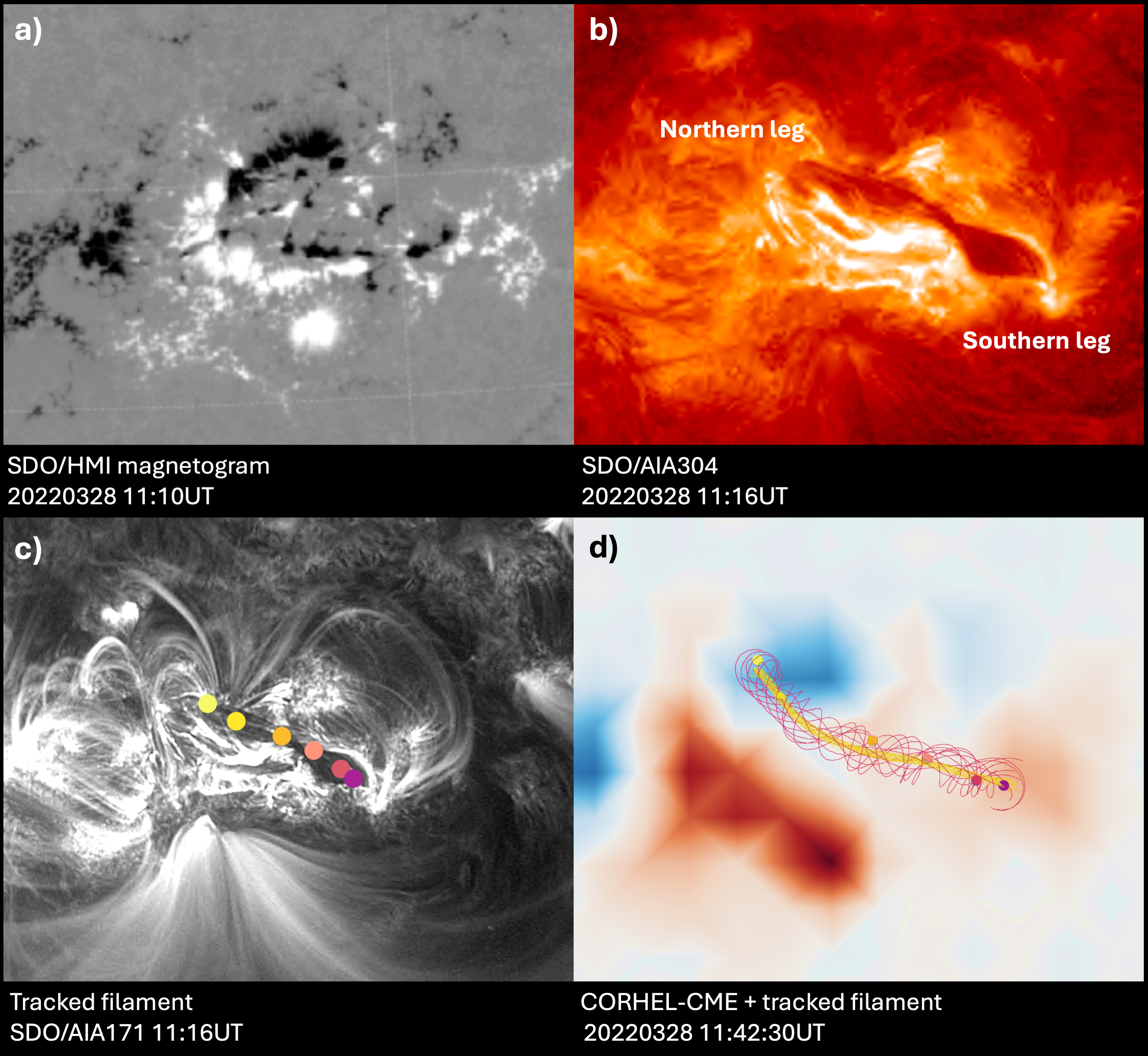}
\caption{a) HMI magnetogram of the AR 12975 at $\sim$11:10$\,$UT. b) EUV observation from AIA-304 of the filament at $\sim$11:16$\,$UT. c) Initial tracking of the filament performed in \citet{sahade_2024} and used as initial condition for CORHEL-CME run. d) Yellow axes show the initial location of the eruptive filament in the CORHEL-CME simulation. Colored dots represent the tracked filament. \label{fig:Filament}}
\end{figure*}

\begin{figure*}[ht!]
\epsscale{1.10}
\begin{interactive}{animation}{20220328_Source.mp4}
\plotone{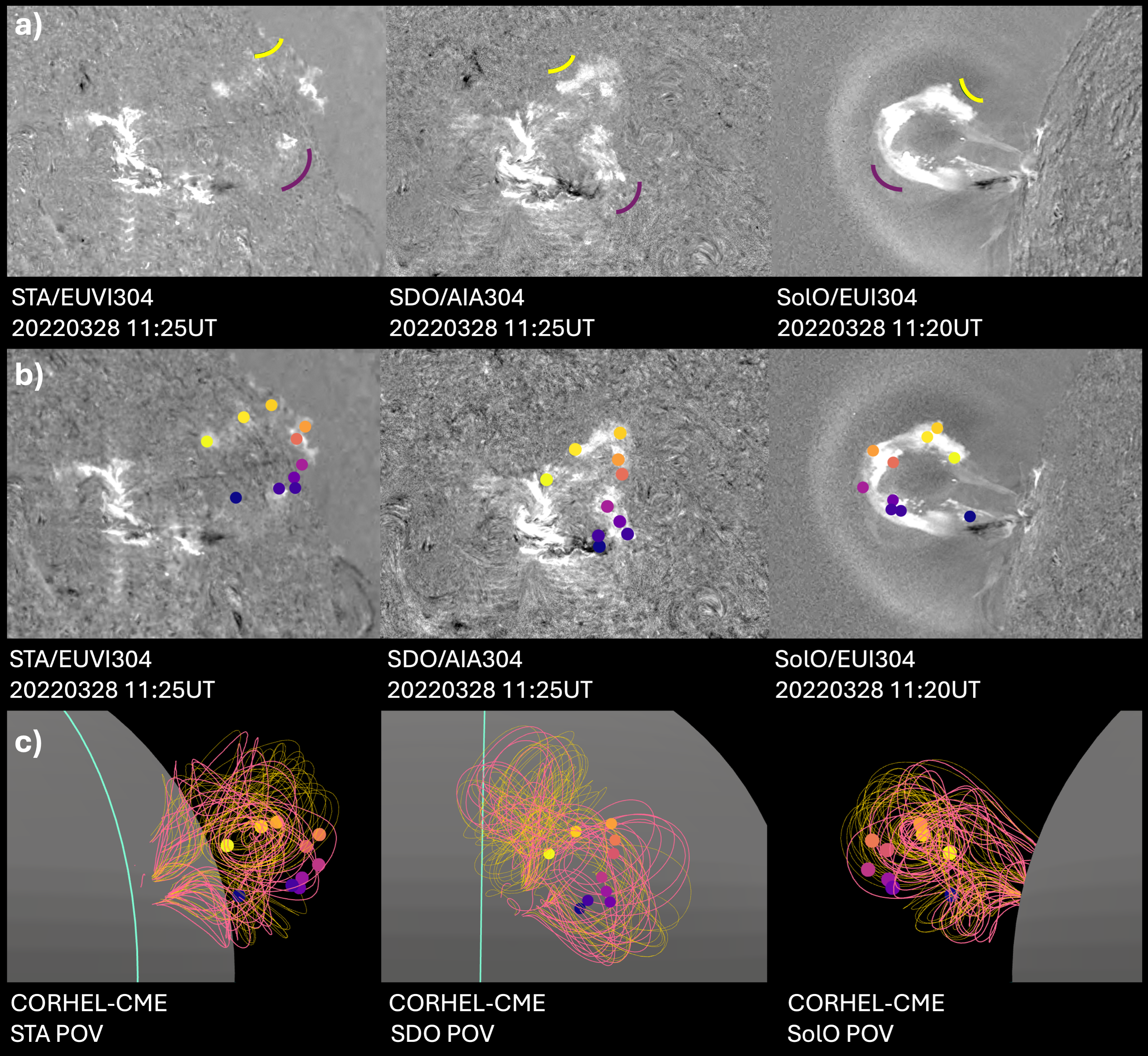}%{EUV_track.png}%{EUV_eruption.png}
\end{interactive}
\caption{a) Running-difference snapshots of the early stage of the eruption in the AIA-304 channel (middle panel), EUVI-304 (left panel) and EUI-304 (right panel). b) Same as panel a but with the results for the tracked filament over-plotted in the three 304\AA channels. Yellow-orange dots represent the northern portion of the filament and Blue-violet dots represents the southern portion (northern and southern legs in Figure \ref{fig:Filament}, respectively). The colors are used to  represent common structures in the various POVs. c) comparison of the tracked filament and the CORHEL-CME model at 11:42:30UT. {An animated version of the filament eruption and early evolution of the CME from the three POVs is available for this figure. The linked movie starts with the filament eruption from AIA-304 POV, in second 5 we show the EUVI-304 POV and in second 8 the EUI-304 POV movie.} \label{fig:EUV_obs}}
\end{figure*}

To understand the evolution{ of the erupting prominence} and its trajectory, \citet{sahade_2024} tracked different features in the EUV observations. The source region and filament were detected by the following EUV instruments: the Atmospheric Imaging Assembly \citep[AIA,][]{lemen_2012} on board Solar Dynamics Observatory \citep[SDO,][]{pesnell_2012}, (hereafter AIA), the Extreme Ultra-Violet Imager \citep[EUVI,][]{wuelser_2004} on board STEREO-A (hereafter EUVI), and Extreme Ultraviolet Imager \citep[EUI,][]{rochus_2020} on board SolO (EUI).  We implemented the tie-pointing technique \citep{inhester_2006} adapted to 3 POVs \citep{scc_measure3}. {Given the position of the filament and the configuration of the spacecraft, the complementary POVs allowed a robust tracking of its evolution. The routine is available at \url{https://github.com/asahade/IDL} and the tracking at \url{https://github.com/asahade/Events}.}

{With the understanding of the filament position and its early evolution given by the coverage of the various spacecraft }and the tie-pointing technique, {we use this information to establish} the initial conditions of the thermodynamic MHD model, CORHEL-CME.

{CORHEL-CME allows} the user to simulate full physics-based CMEs in a coronal environment and solar wind derived by synoptic magnetograms. It presents three fundamental steps: 1) Creation of the eruptive MFR with a zero-beta MHD simulation. 2) Thermodynamic MHD solution of the environment, where the corona is reconstructed thermodynamically by selecting a heating model, in which the MFR from step 1 is inserted, 3) Full physics-based CME simulations from $1\,R_\odot$ to 1$\,$au. As an output, the user obtains not only the evolution of magnetic field lines but also the thermodynamic properties and the synthetic EUV emission of the simulated CME. More details can be found in \citet{linker_2024}.
{It is not the goal} of the model to perfectly reproduce the eruption, but to understand the global evolution and external interactions of the {modeled CME. }%This is a powerful tool to interpret the physics behind the distorted large scale structures observed in the white-light images.

Figure \ref{fig:Filament}a shows the magnetogram {of AR 12975 }provided by the Helioseismic Magnetic Imager \citep[HMI]{scherrer_2012} onboard the SDO mission. Figure \ref{fig:Filament}b shows the filament before its eruption as observed by AIA-304. The elongated right-handed filament overlies a complicated polarity inversion line (PIL) and it is oriented along the NE-SW direction, tilted around -20$^{\circ}$ clockwise from the solar equator, with its northern (southern) leg rooted in a negative (positive) polarity patch.

Figure \ref{fig:Filament}c shows the 3D position of tracked features within the filament at around 11:10$\,$UT from the AIA-304 POV. Figure \ref{fig:Filament}d shows the MFR created by the step 1 of CORHEL-CME, which matches the position, height and curvature of the filament (see colored tracked dots for reference). The details of the parameters and zero-beta simulations are available at CCMC simulations results: \href{https://ccmc.gsfc.nasa.gov/results/viewrun.php?runnumber=Abril_Sahade_042424_SH_1}{Abril\_Sahade\_042424\_SH\_1}.

The EUV observations and tracked results indicate that the northern leg (see Figure \ref{fig:Filament}b) erupted faster than the southern one. This asymmetric eruption leads to rotational motions that dragged the eruptive material to a N-S orientation ($-90^\circ$ tilt), as the 3D tracking confirmed. Also, the filament was deflected towards the west coronal hole, and interacted with it. 

Figure \ref{fig:EUV_obs}a shows an early stage of the eruption in AIA-304 (middle panel), EUVI-304 (left panel), and EUI-304 (right panel). The yellow and violet curves mark the same structures from the different POV. Since SolO was located at 0.3$\,$au, the observation times were selected to match the same moments as the observations. Figure \ref{fig:EUV_obs}b shows an example of the tracked dots projected over the EUV observations. 

{We validate the parameters selected for step 1 of the CORHEL-CME model by matching the shape and location of the tracked filament with the simulated MFR.} While the initial position was determined by the pre-eruptive filament, the fraction of optimized current was chosen to match the rotation and deflection of the eruption. Figure \ref{fig:EUV_obs}c shows an over-plot of the MFR field lines of the CORHEL-CME run and the tracked filament (EUVI, AIA, and EUI on left, middle and right panel, respectively). It can be seen that the tracked filament is `embedded' within the simulated MFR.
{To compare the simulation with observations, we matched the height of the MFR with the filament and, eventually, the CME. This approach ensures that the spatial evolution of the eruption aligns between the model and the data, even if the timing does not precisely coincide.
The simulated eruption initially evolves faster than the observed one but slows down at later stages. This discrepancy likely reflects missing physical processes in the model, such as flux emergence or flare reconnection, which may influence the eruption.
Despite these limitations and the discrepancies between the kinematics of the real and simulated CME, matching the spatial position throughout the evolution gives us confidence that the simulation is capturing key features of the event. In particular, it reproduces the rapid rise of the filament’s northern leg which pulls the southern portion upward.} %—as well as the three distinct structures identified in white-light images: The U-shape concave structure denoted by (curve 1), the halo CME (curve 2) and the v-shaped convex front observed in SoloHI FOV (curve 3).

The full results of the CORHEL-CME run, including the magnetic field evolution and synthetic emission of the eruption, can be found at the CCMC simulations results page: \href{https://ccmc.gsfc.nasa.gov/results/viewrun.php?runnumber=Abril_Sahade_052324_SH_2}{Abril\_Sahade\_052324\_SH\_2}.

Based on the observational analysis, we believe that the March 28 CME was generated by a single filament-MFR eruption that underwent rotational motions and interacted with an overlying helmet streamer. However, it is not clear why the resulting CME departs so much from the typical MFR morphology, presenting signatures of two differently oriented cavities. 

To clarify how this system evolved through the corona and what is correlating with the complex structures observed in white-light images (see Section \ref{sec:data}), we investigate the results of the CORHEL-CME model for the March 28 event in the next section.

\section{Results}%{Comparison with the CORHEL-CME model}
\label{sec:Model}

\begin{figure*}[ht!]
\epsscale{1.15}
\plotone{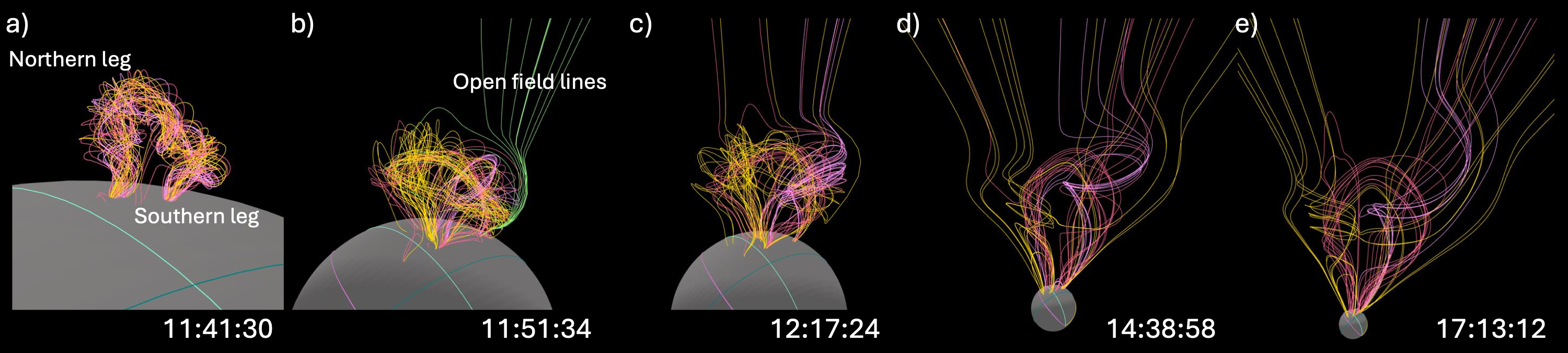}
\caption{Snapshots of the evolution of the MFR obtained by CORHEL-CME from an arbitrary POV. The gray sphere ($1\,R_\odot$) has the Earth meridian (green line), equator (teal line), and STA meridian (pink line) plotted for reference. Panel a) early rise of the MFR showing the northern leg faster ascent. Panel b) the southern leg started to reconnect with the open field lines of the west coronal hole (green lines). Panel c) Both legs of the MFR reached a similar height, the northern leg started to reconnect with the east coronal hole after escaping the helmet streamer arcade. Panel d) The MFR reaches a balance with the ambient field, evolving mostly self-similarly. The inner MFR (pink lines) remains attached to the solar surface at both ends. Panel e) later snapshot of the evolution showing similar characteristics of previous panel after more expansion (see the size of the sphere for reference).
\label{fig:CORHEL}}
\end{figure*}

\begin{figure*}[ht!]
\epsscale{0.9}
\begin{interactive}{js}{3d_CORHEL-CME_a.html.zip}
\plotone{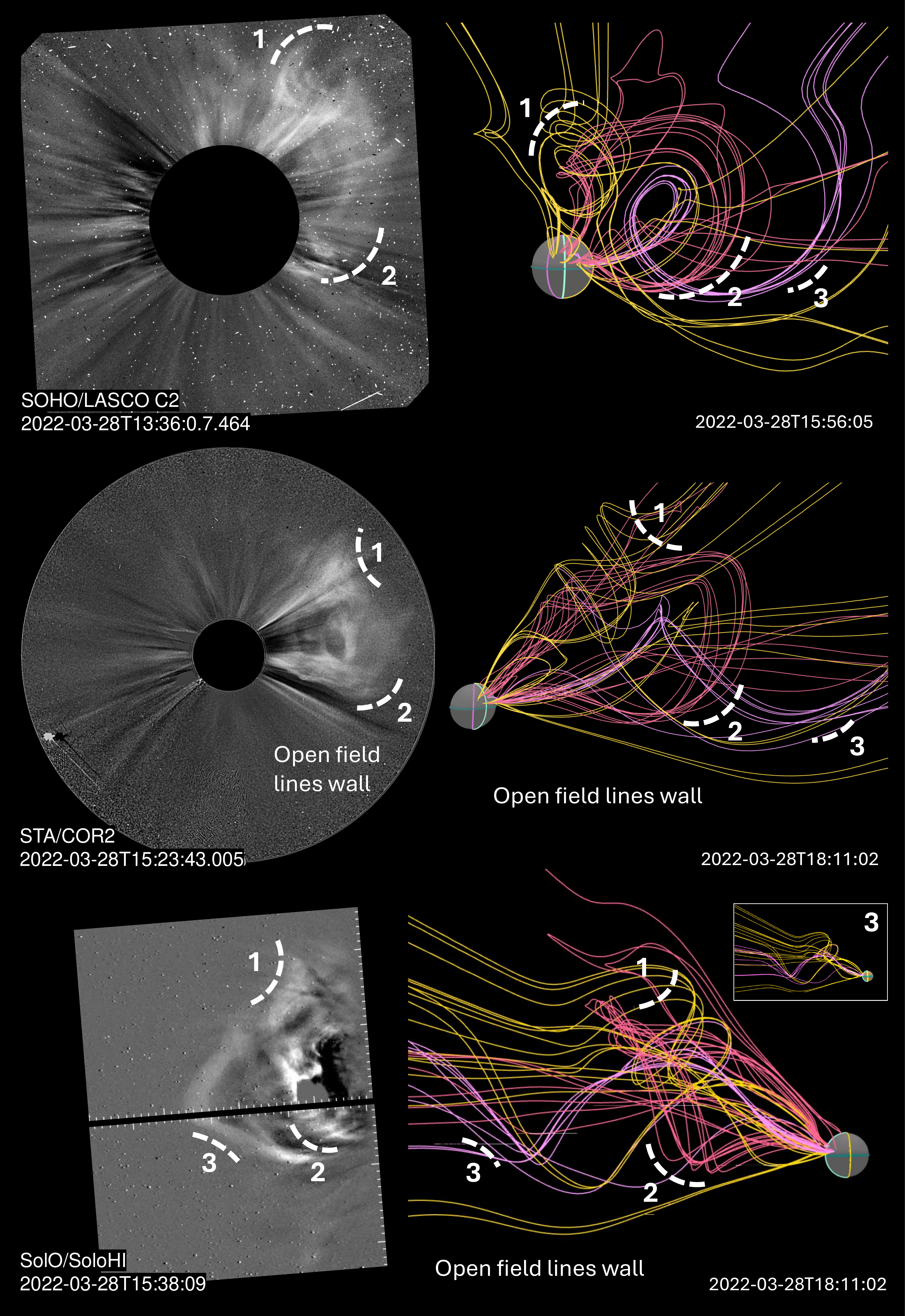}%{CORHEL-CME.pdf}%Comparison-CORHEL.pdf}
\end{interactive}
\caption{Left panel: white-light observations provided by LASCO (upper panel), COR2 (middle panel) and SoloHI (Lower panel) for the March 28 event. Right panel: comparison with the CORHEL-CME model from the perspective of each specific spacecraft. {The gray sphere ($1\,R_\odot$) has the Earth meridian (light-green line), STA meridian (pink line) and SolO meridian (yellow line), plotted for reference. The last two are at $-33.3^{\circ}$ and $83.9^{\circ}$ from the Earth meridian, respectively.} In both cases white curves denote the same structures identified in the model and the observations. The cutout on the bottom-right panel provides a mini-view of the simulation form SoloHI POV where the MFR core of is removed. This reveals a field line configuration similar to a v-shape denoted with curve 3. {An interactive view of the right panels of this figure displayed together on a 3D visualization is available online. The interactive view allows users to hover, zoom and rotate the figure to explore it in a more comprehensive way.}
\label{fig:CompCORHEL}}
\end{figure*}

%The magnetic field evolution from 1 to 30$\,R_\odot$ obtained by CORHEL-CME (available \href{https://ccmc.gsfc.nasa.gov/RoR_WWW/SH/CORHEL-CME/2024/Abril_Sahade_052324_SH_2/Abril_Sahade_052324_SH_2_cme_thermocme_report/mapfl_sequence/traces.html}{here}) can help in connecting and interpreting the EUV and white-light observations. 

{Figure \ref{fig:CORHEL} }shows snapshots of the MFR evolution from an arbitrary POV. The field lines plotted represent the inner MFR (pink lines) and some lines that originally belonged to the MFR and reconnected with the magnetic environment (yellow and purple lines). Most of the yellow field lines indicate the evolution of the northern portion of the MFR, while the purple field lines are connected to the southern portion.

{As observed }in the cold plasma of the {erupting prominence, the northern portion} of the MFR reaches greater heights than the southern one in its early evolution (panel a in the figure), rotating the system and increasing its tilt angle. {This may be due }to either an internal imbalance of the regularized Biot-Savart flux-rope \citep{titov_2018}, its interaction with an inhomogeneous strapping field above each section or, more likely, a combination of both factors. Afterwards, it is the outer envelope of the southern MFR portion that begins to reconnect with the west coronal hole (panel b). This reconnection removes strapping field over the southern MFR portion allowing it to reach a height comparable to the northern portion, which after the first faster rise is slowed by the closed field of the helmet streamer arcade (panel c). Later on, the northern portion of the MFR finally escapes reaching the east coronal hole. Due to the different environment in which each portion evolved (closed vs. open field), the MFR expanded differently at each end. The southern portion expanded more and the reconnected front is smoothly shaped and pushed by the west coronal hole. On the other hand, the northern portion shows less expansion and a more compact circular envelope created by the piling up of the closed arcade (panel c and d of Figure \ref{fig:CORHEL}). The results obtained by the simulation can be compared with the white-light structures in order to understand their magnetic nature.

Figure \ref{fig:CompCORHEL} shows a comparison between the model and a comparable time of the CME for each instrument. The left panels display the observations provided by LASCO C2 (upper panel), COR2 (middle panel), and SoloHI (lower panel). The right panel shows the MHD model, rotated to the perspective of each specific spacecraft. White curves denote the different structures that can be identified in the model and the observations. 

{It is} possible to see in the Figure, and in the full simulation, that most of the yellow field lines (northern portion of the MFR) preserve most of the MFR twist (also seen in the inner MFR). The MFR southern portion (purple field lines) started their reconnection with the coronal hole earlier in time (about 12 UT in the simulation), showing a smoother shape and transferring the {front edge} away faster which separated it from the inner MFR (pink lines).

The upper panels show the comparison with LASCO, where the cavity structure and rings observed (denoted by curve 1) are reproduced by the northern MFR portion of the MHD model (most yellow field lines). On the other hand, the halo (with curve 2 as its lower boundary) matches the orientation and appearance of the inner MFR (pink lines).
From another perspective, the middle panels show the comparison between COR2 and the model. All the features described above for LASCO C2 POV can also be seen from this viewpoint. The upper yellow lines related to the northern part of the filament are clearly entangled with the inner MFR (pink lines) generating the U-shaped structure denoted by curve 1. {Moreover, the whole inner MFR, orientated slightly to the north, resulted in the leading edge highlighted {with curve 2}. The back portion of the magnetic field lines are oriented to produce the cavity signatures from this POV. Purple lines form a front that evolves ahead the inner MFR (curve 3) but it cannot be identified in the LASCO C2 and COR2 FOV.}

The lower panel shows the comparison with SoloHI observations. As mentioned before, the concave distorted front is correlated with the {entangled} magnetic field that result from the northern portion of the MFR evolution, which from SoloHI perspective looks messier and produces different density enhancements as the distorted portion highlighted {with curve 1}. Once again, the inner MFR (pink lines) produces the more regular leading edge that from this POV resulted in a more flattened front (curve 2).

The combination of the southern portion (purple lines), shaped by the west coronal hole, and the upper portion (yellow lines), driven by the helmet streamer, sets the stage for {the formation of a v-shaped front edge} in a radial direction rather than northwards. The bottom panel of Figure \ref{fig:CompCORHEL} provides an alternative mini-view, where the core of the MFR is removed, revealing a field line configuration similar to a v-shape (curve 3). More external magnetic field lines, not shown here, may pile up at the front, resulting in the density enhancement observed in SoloHI FOV.

{We do not claim that the modeled field lines or the resulting structures are exact reconstructions of the observed event, but rather we believe they can provide a general understanding of how and when these diverse structures could be generated to help explain the obvservations. The CORHEL-CME model illustrates how the reconnection and internal evolution of a single MFR can create the observed cavity (curve 3 and yellow field lines), along with a more typical CME morphology (curve 2 and pink lines).}

\section{Discussion and Conclusions}
\label{sec:conc}

CMEs are complicated phenomena that seldom{ match simplified geometrical models.} As CMEs evolve from the solar corona into the heliosphere, their internal structure changes. Through the analysis of continuous data provided by 1$\,$au missions, we gained understanding of the global structure of CMEs by simplifying their main characteristics. Today, with the superior effective resolution and alternative viewpoints offered by encounter missions, {we are equipped} to reveal detailed structures and complexities of CMEs, refining our understanding. At the same time, interpretations of encounter missions such as Solar Orbiter are significantly enhanced by synoptic missions, {which provide the broader context of CME evolution}. A multi-perspective approach, complemented with higher-resolution observations and sophisticated models, is crucial for deciphering CME structure. 

Using a careful analysis of the source region and its environment, we modeled the 28 March event with the CORHEL-CME model. By comparing the {observational data} with the modeled magnetic configuration, we deduce that the atypical CME morphology was the result of the MFR rotation and its interaction with the ambient field.
Feature 1 in Figure \ref{fig:LC2_0328} is the result of the MFR northern portion that underwent a complex evolution after an initial destabilization and{ fast rise }obstructed by the closed overlying field belonging to the helmet streamer. That caused distortion and entanglement of the magnetic field lines (see yellow and upper portion of pink field lines in Figure \ref{fig:CompCORHEL}). As a result, the structure appeared, in white-light images, as the bottom part of an erupting MFR, located above the expected CME front (indicated by curve 2), without being an actual MFR. Only through the comparison between modeling and observations were we able to reach this conclusion and resolve the confusing white-light signatures of this event. This leads to an important lesson and a word of caution. White-light morphologies, such as U-shape arcs, that have long been considered as tell-tale MFR signatures, should not be taken at face-value, especially when the overall appearance does not conform to other signatures. This was the case for this event---a single eruption that resulted {in two apparent MFRs }in the coronagraph images.

However, the southern portion of the MFR evolved along a lower resistance path towards the open field lines with positive polarity. The southern MFR portion reconnected and evolved smoothly producing a more traditional front and cavity morphology. The angularly-shaped {front} ahead of the CME, seen only by SoloHI, was another curiosity solved by the comparison to the MHD model. As can be seen in Figure~\ref{fig:CompCORHEL} (bottom right), the {v-shaped front} reflects the orientation of the upstream field lines.  

The CORHEL-CME model was key for interpreting the unusual white-light structures, helping to clarify the relationship between the different fronts and {their formation mechanisms by} investigating their magnetic origin. The success underscores the value of combining high-resolution observations with physics-based modeling to enhance our understanding of CME evolution. The synergy between observations and modeling {offers a powerful path forward for advancing our knowledge of CME interactions and their implications for space weather, providing a strong foundation for future research.}

\begin{acknowledgments}

C.M. is supported by the NASA Heliophysics Division, Solar Orbiter Collaboration Office under DPR NNG09EK11I. A.S. is supported by an appointment to the NASA Postdoctoral Program at the NASA Goddard Space Flight Center, administered by Oak Ridge Associated Universities under contract with NASA. A.V. is supported by NASA grant 80NSSC22K1028 and 80NSSC22K0970. P.H. and R.C. are supported by the Office of Naval Research. T.N.-C. thanks the support of the Solar Orbiter mission. Simulation results have been provided by the Community Coordinated Modeling Center (CCMC) at Goddard Space Flight Center through their publicly available simulation services (https://ccmc.gsfc.nasa.gov). The CORHEL-CME Model was developed by Jon Linker at the Predictive Science Inc. The authors would like to thank Martin Reiss for his willingness and assistance in implementing CORHEL-CME. 
Solar Orbiter is a mission of international cooperation between the European Space Agency (ESA) and the National Aeronautics and Space Administration (NASA), operated by ESA. The Solar Orbiter Heliospheric Imager (SoloHI) instrument was designed, built, and is now operated by the US Naval Research Laboratory with the support of the NASA Heliophysics Division, Solar Orbiter Collaboration Office under DPR NNG09EK11I. 

\end{acknowledgments}

\begin{contribution}
%%This section gives authors the space to recognize author contributions. The text inside this environment is NOT counted towards the total word quanta. At a minimum, manuscripts are expected to include this text:

CM started the initial concept of the research, processed the images, contributed in the analysis and wrote the manuscript. AS contributed MHD modeling expertise, conducted CME tracking, and executed CORHEL-CME simulations. She also collaborated closely with the PSI and CCMC teams to enhance the outputs and played a key role in the analysis and writing.  AV played a key role in interpreting the results and refining the manuscript. PH, RC, and TNC assisted with analysis and edited the manuscript.

%% But authors are expected to provide more specific details, e.g. 
%%
%%SC was responsible for writing and submitting the manuscript.
%%WWM came up with the initial research concept and edited the manuscript.
%%OTS obtained the funding and edited the manuscript.
%%EBF provided the formal analysis and validation. He also edited the manuscript.
%%GEH Supervised the undergraduates, wrote the software and administers the project github and Zenodo repositories.
%%
%% Authors can use the Contributor Role Taxonomy (CRediT) at
%% for ideas on how write a good statement tailored to their needs.

\end{contribution}

\facilities{SolO/EUI, SDO/AIA, STEREO-A/EUVI, SOHO/LASCO-C2, STEREO/COR2 and SolO/SoloHI data}

%\software{astropy \citep{2013A&A...558A..33A,2018AJ....156..123A,2022ApJ...935..167A},  
%          Cloudy \citep{2013RMxAA..49..137F}, 
%          Source Extractor \citep{1996A&AS..117..393B}
%          }


%% For this sample we use BibTeX plus aasjournalv7.bst to generate the

\bibliography{sample7}{}

\begin{thebibliography}{}
\expandafter\ifx\csname natexlab\endcsname\relax\def\natexlab#1{#1}\fi
\providecommand{\url}[1]{\href{#1}{#1}}
\providecommand{\dodoi}[1]{doi:~\href{http://doi.org/#1}{\nolinkurl{#1}}}
\providecommand{\doeprint}[1]{\href{http://ascl.net/#1}{\nolinkurl{http://ascl.net/#1}}}
\providecommand{\doarXiv}[1]{\href{https://arxiv.org/abs/#1}{\nolinkurl{https://arxiv.org/abs/#1}}}

\bibitem[{G.~E. {Brueckner} {et~al.}(1995){Brueckner}, {Howard}, {Koomen}, {Korendyke}, {Michels}, {Moses}, {Socker}, {Dere}, {Lamy}, {Llebaria}, {Bout}, {Schwenn}, {Simnett}, {Bedford}, \& {Eyles}}]{brueckner_1995}
{Brueckner}, G.~E., {Howard}, R.~A., {Koomen}, M.~J., {et~al.} 1995, \bibinfo{title}{{The Large Angle Spectroscopic Coronagraph (LASCO)},} \solphys, 162, 357, \dodoi{10.1007/BF00733434}

\bibitem[{P.~J. {Cargill} \& J.~M. {Schmidt}(2002){Cargill} \& {Schmidt}}]{cargill_2002}
{Cargill}, P.~J., \& {Schmidt}, J.~M. 2002, \bibinfo{title}{{Modelling interplanetary CMEs using magnetohydrodynamic simulations},} Annales Geophysicae, 20, 879, \dodoi{10.5194/angeo-20-879-2002}

\bibitem[{R.~C. {Colaninno} \& A. {Vourlidas}(2015){Colaninno} \& {Vourlidas}}]{colaninno_2015}
{Colaninno}, R.~C., \& {Vourlidas}, A. 2015, \bibinfo{title}{{Using Multiple-viewpoint Observations to Determine the Interaction of Three Coronal Mass Ejections Observed on 2012 March 5},} \apj, 815, 70, \dodoi{10.1088/0004-637X/815/1/70}

\bibitem[{V. {Domingo} {et~al.}(1995){Domingo}, {Fleck}, \& {Poland}}]{domingo_1995}
{Domingo}, V., {Fleck}, B., \& {Poland}, A.~I. 1995, \bibinfo{title}{{The SOHO Mission: an Overview},} \solphys, 162, 1, \dodoi{10.1007/BF00733425}

\bibitem[{C. {Downs} {et~al.}(2013){Downs}, {Linker}, {Miki{\'c}}, {Riley}, {Schrijver}, \& {Saint-Hilaire}}]{downs_2013}
{Downs}, C., {Linker}, J.~A., {Miki{\'c}}, Z., {et~al.} 2013, \bibinfo{title}{{Probing the Solar Magnetic Field with a Sun-Grazing Comet},} Science, 340, 1196, \dodoi{10.1126/science.1236550}

\bibitem[{P. {Hess} {et~al.}(2023){Hess}, {Colaninno}, {Vourlidas}, {Howard}, \& {Stenborg}}]{hess_2023}
{Hess}, P., {Colaninno}, R.~C., {Vourlidas}, A., {Howard}, R.~A., \& {Stenborg}, G. 2023, \bibinfo{title}{{SoloHI observations of coronal mass ejections observed by multiple spacecraft},} \aap, 679, A149, \dodoi{10.1051/0004-6361/202346907}

\bibitem[{R.~A. {Howard} {et~al.}(2008){Howard}, {Moses}, {Vourlidas}, {Newmark}, {Socker}, {Plunkett}, {Korendyke}, {Cook}, {Hurley}, {Davila}, \& {et al.}}]{howard_2008}
{Howard}, R.~A., {Moses}, J.~D., {Vourlidas}, A., {et~al.} 2008, \bibinfo{title}{{Sun Earth Connection Coronal and Heliospheric Investigation (SECCHI)},} \ssr, 136, 67, \dodoi{10.1007/s11214-008-9341-4}

\bibitem[{R.~A. {Howard} {et~al.}(2020){Howard}, {Vourlidas}, {Colaninno}, {Korendyke}, {Plunkett}, {Carter}, {Wang}, {Rich}, {Lynch}, {Thurn}, {Socker}, {Thernisien}, {Chua}, {Linton}, {Koss}, {Tun-Beltran}, {Dennison}, {Stenborg}, {McMullin}, {Hunt}, {Baugh}, {Clifford}, {Keller}, {Janesick}, {Tower}, {Grygon}, {Farkas}, {Hagood}, {Eisenhauer}, {Uhl}, {Yerushalmi}, {Smith}, {Liewer}, {Velli}, {Linker}, {Bothmer}, {Rochus}, {Halain}, {Lamy}, {Auch{\`e}re}, {Harrison}, {Rouillard}, {Patsourakos}, {St. Cyr}, {Gilbert}, {Maldonado}, {Mariano}, \& {Cerullo}}]{howard_2020}
{Howard}, R.~A., {Vourlidas}, A., {Colaninno}, R.~C., {et~al.} 2020, \bibinfo{title}{{The Solar Orbiter Heliospheric Imager (SoloHI)},} \aap, 642, A13, \dodoi{10.1051/0004-6361/201935202}

\bibitem[{J. {Hutton} \& H. {Morgan}(2015){Hutton} \& {Morgan}}]{hutton_2015}
{Hutton}, J., \& {Morgan}, H. 2015, \bibinfo{title}{{Erupting Filaments with Large Enclosing Flux Tubes as Sources of High-mass Three-part CMEs, and Erupting Filaments in the Absence of Enclosing Flux Tubes as Sources of Low-mass Unstructured CMEs},} \apj, 813, 35, \dodoi{10.1088/0004-637X/813/1/35}

\bibitem[{R.~M.~E. {Illing} \& A.~J. {Hundhausen}(1985){Illing} \& {Hundhausen}}]{illing_1985}
{Illing}, R.~M.~E., \& {Hundhausen}, A.~J. 1985, \bibinfo{title}{{Observation of a coronal transient from 1.2 to 6 solar radii},} \jgr, 90, 275, \dodoi{10.1029/JA090iA01p00275}

\bibitem[{B. {Inhester}(2006){Inhester}}]{inhester_2006}
{Inhester}, B. 2006, \bibinfo{title}{{Stereoscopy basics for the STEREO mission},} arXiv e-prints, astro, \dodoi{10.48550/arXiv.astro-ph/0612649}

\bibitem[{M.~L. {Kaiser} {et~al.}(2008){Kaiser}, {Kucera}, {Davila}, {St.~Cyr}, {Guhathakurta}, \& {Christian}}]{kaiser_2008}
{Kaiser}, M.~L., {Kucera}, T.~A., {Davila}, J.~M., {et~al.} 2008, \bibinfo{title}{{The STEREO mission: An introduction},} Space Sci. Rev., 136, 5, \dodoi{10.1007/s11214-007-9277-0}

\bibitem[{C. {Kay} {et~al.}(2017){Kay}, {Gopalswamy}, {Xie}, \& {Yashiro}}]{kay_2017}
{Kay}, C., {Gopalswamy}, N., {Xie}, H., \& {Yashiro}, S. 2017, \bibinfo{title}{{Deflection and Rotation of CMEs from Active Region 11158},} \solphys, 292, 78, \dodoi{10.1007/s11207-017-1098-z}

\bibitem[{J.~R. {Lemen} {et~al.}(2012){Lemen}, {Title}, {Akin}, {Boerner}, {Chou}, {Drake}, {Duncan}, {Edwards}, {Friedlaender}, {Heyman}, \& {et al.}}]{lemen_2012}
{Lemen}, J.~R., {Title}, A.~M., {Akin}, D.~J., {et~al.} 2012, \bibinfo{title}{{The Atmospheric Imaging Assembly (AIA) on the Solar Dynamics Observatory (SDO)},} \solphys, 275, 17, \dodoi{10.1007/s11207-011-9776-8}

\bibitem[{J.~A. {Linker} {et~al.}(2024){Linker}, {Torok}, {Downs}, {Caplan}, {Titov}, {Reyes}, {Lionello}, \& {Riley}}]{linker_2024}
{Linker}, J.~A., {Torok}, T., {Downs}, C., {et~al.} 2024, in Journal of Physics Conference Series, Vol. 2742, Journal of Physics Conference Series (IOP), 012012, \dodoi{10.1088/1742-6596/2742/1/012012}

\bibitem[{R. {Lionello} {et~al.}(2009){Lionello}, {Linker}, \& {Miki{\'c}}}]{lionello_2009}
{Lionello}, R., {Linker}, J.~A., \& {Miki{\'c}}, Z. 2009, \bibinfo{title}{{Multispectral Emission of the Sun During the First Whole Sun Month: Magnetohydrodynamic Simulations},} \apj, 690, 902, \dodoi{10.1088/0004-637X/690/1/902}

\bibitem[{C. {Mac Cormack} {et~al.}(2025){Mac Cormack}, {Shaik}, {Hess}, {Colaninno}, \& {Nieves-Chinchilla}}]{maccormack_2025}
{Mac Cormack}, C., {Shaik}, S.~B., {Hess}, P., {Colaninno}, R., \& {Nieves-Chinchilla}, T. 2025, \bibinfo{title}{{A Multi-Viewpoint CME Catalog Based on SoloHI Observed Events},} \solphys, 300, 54, \dodoi{10.1007/s11207-025-02463-7}

\bibitem[{D.~E. {Morosan} {et~al.}(2024){Morosan}, {Pomoell}, {Palmroos}, {Dresing}, {Asvestari}, {Vainio}, {Kilpua}, {Gieseler}, {Kumari}, \& {Jebaraj}}]{morosan_2024}
{Morosan}, D.~E., {Pomoell}, J., {Palmroos}, C., {et~al.} 2024, \bibinfo{title}{{Connecting remote and in situ observations of shock-accelerated electrons associated with a coronal mass ejection},} \aap, 683, A31, \dodoi{10.1051/0004-6361/202347873}

\bibitem[{D. {M{\"u}ller} {et~al.}(2020){M{\"u}ller}, {St. Cyr}, {Zouganelis}, {Gilbert}, {Marsden}, {Nieves-Chinchilla}, {Antonucci}, {Auch{\`e}re}, {Berghmans}, {Horbury}, {Howard}, {Krucker}, {Maksimovic}, {Owen}, {Rochus}, {Rodriguez-Pacheco}, {Romoli}, {Solanki}, {Bruno}, {Carlsson}, {Fludra}, {Harra}, {Hassler}, {Livi}, {Louarn}, {Peter}, {Sch{\"u}hle}, {Teriaca}, {del Toro Iniesta}, {Wimmer-Schweingruber}, {Marsch}, {Velli}, {De Groof}, {Walsh}, \& {Williams}}]{muller_2020}
{M{\"u}ller}, D., {St. Cyr}, O.~C., {Zouganelis}, I., {et~al.} 2020, \bibinfo{title}{{The Solar Orbiter mission. Science overview},} \aap, 642, A1, \dodoi{10.1051/0004-6361/202038467}

\bibitem[{S. {Patsourakos} {et~al.}(2020){Patsourakos}, {Vourlidas}, {T{\"o}r{\"o}k}, {Kliem}, {Antiochos}, {Archontis}, {Aulanier}, {Cheng}, {Chintzoglou}, {Georgoulis}, {Green}, {Leake}, {Moore}, {Nindos}, {Syntelis}, {Yardley}, {Yurchyshyn}, \& {Zhang}}]{patsourakos_2020}
{Patsourakos}, S., {Vourlidas}, A., {T{\"o}r{\"o}k}, T., {et~al.} 2020, \bibinfo{title}{{Decoding the Pre-Eruptive Magnetic Field Configurations of Coronal Mass Ejections},} \ssr, 216, 131, \dodoi{10.1007/s11214-020-00757-9}

\bibitem[{W.~D. {Pesnell} {et~al.}(2012){Pesnell}, {Thompson}, \& {Chamberlin}}]{pesnell_2012}
{Pesnell}, W.~D., {Thompson}, B.~J., \& {Chamberlin}, P.~C. 2012, \bibinfo{title}{{The Solar Dynamics Observatory (SDO)},} \solphys, 275, 3, \dodoi{10.1007/s11207-011-9841-3}

\bibitem[{T. {Podladchikova} {et~al.}(2024){Podladchikova}, {Jain}, {Veronig}, {Purkhart}, {Chikunova}, {Dissauer}, \& {Dumbovi{\'c}}}]{podladchikova_2024}
{Podladchikova}, T., {Jain}, S., {Veronig}, A.~M., {et~al.} 2024, \bibinfo{title}{{Three-part structure of a solar coronal mass ejection observed in low coronal signatures of Solar Orbiter},} \aap, 691, A344, \dodoi{10.1051/0004-6361/202451777}

\bibitem[{S. {Purkhart} {et~al.}(2023){Purkhart}, {Veronig}, {Dickson}, {Battaglia}, {Krucker}, {Jarolim}, {Kliem}, {Dissauer}, {Podladchikova}, {STIX Team}, \& {EUI Team}}]{purkhart_2023}
{Purkhart}, S., {Veronig}, A.~M., {Dickson}, E. C.~M., {et~al.} 2023, \bibinfo{title}{{Multipoint study of the energy release and transport in the 28 March 2022, M4 flare using STIX, EUI, and AIA during the first Solar Orbiter nominal mission perihelion},} \aap, 679, A99, \dodoi{10.1051/0004-6361/202346354}

\bibitem[{P. {Riley} {et~al.}(2003){Riley}, {Linker}, {Miki{\'c}}, {Odstrcil}, {Zurbuchen}, {Lario}, \& {Lepping}}]{riley_2003}
{Riley}, P., {Linker}, J.~A., {Miki{\'c}}, Z., {et~al.} 2003, \bibinfo{title}{{Using an MHD simulation to interpret the global context of a coronal mass ejection observed by two spacecraft},} Journal of Geophysical Research (Space Physics), 108, 1272, \dodoi{10.1029/2002JA009760}

\bibitem[{P. {Rochus} {et~al.}(2020){Rochus}, {Auch{\`e}re}, {Berghmans}, {Harra}, {Schmutz}, {Sch{\"u}hle}, {Addison}, {Appourchaux}, {Aznar Cuadrado}, {Baker}, {Barbay}, {Bates}, {BenMoussa}, {Bergmann}, {Beurthe}, {Borgo}, {Bonte}, {Bouzit}, {Bradley}, {B{\"u}chel}, {Buchlin}, {B{\"u}chner}, {Cab{\'e}}, {Cadiergues}, {Chaigneau}, {Chares}, {Choque Cortez}, {Coker}, {Condamin}, {Coumar}, {Curdt}, {Cutler}, {Davies}, {Davison}, {Defise}, {Del Zanna}, {Delmotte}, {Delouille}, {Dolla}, {Dumesnil}, {D{\"u}rig}, {Enge}, {Fran{\c{c}}ois}, {Fourmond}, {Gillis}, {Giordanengo}, {Gissot}, {Green}, {Guerreiro}, {Guilbaud}, {Gyo}, {Haberreiter}, {Hafiz}, {Hailey}, {Halain}, {Hansotte}, {Hecquet}, {Heerlein}, {Hellin}, {Hemsley}, {Hermans}, {Hervier}, {Hochedez}, {Houbrechts}, {Ihsan}, {Jacques}, {J{\'e}r{\^o}me}, {Jones}, {Kahle}, {Kennedy}, {Klaproth}, {Kolleck}, {Koller}, {Kotsialos}, {Kraaikamp}, {Langer}, {Lawrenson}, {Le Clech'}, {Lenaerts}, {Liebecq}, {Linder}, {Long}, {Mampaey}, {Markiewicz-Innes}, {Marquet},
  {Marsch}, {Matthews}, {Mazy}, {Mazzoli}, {Meining}, {Meltchakov}, {Mercier}, {Meyer}, {Monecke}, {Monfort}, {Morinaud}, {Moron}, {Mountney}, {M{\"u}ller}, {Nicula}, {Parenti}, {Peter}, {Pfiffner}, {Philippon}, {Phillips}, {Plesseria}, {Pylyser}, {Rabecki}, {Ravet-Krill}, {Rebellato}, {Renotte}, {Rodriguez}, {Roose}, {Rosin}, {Rossi}, {Roth}, {Rouesnel}, {Roulliay}, {Rousseau}, {Ruane}, {Scanlan}, {Schlatter}, {Seaton}, {Silliman}, {Smit}, {Smith}, {Solanki}, {Spescha}, {Spencer}, {Stegen}, {Stockman}, {Szwec}, {Tamiatto}, {Tandy}, {Teriaca}, {Theobald}, {Tychon}, {van Driel-Gesztelyi}, {Verbeeck}, {Vial}, {Werner}, {West}, {Westwood}, {Wiegelmann}, {Willis}, {Winter}, {Zerr}, {Zhang}, \& {Zhukov}}]{rochus_2020}
{Rochus}, P., {Auch{\`e}re}, F., {Berghmans}, D., {et~al.} 2020, \bibinfo{title}{{The Solar Orbiter EUI instrument: The Extreme Ultraviolet Imager},} \aap, 642, A8, \dodoi{10.1051/0004-6361/201936663}

\bibitem[{L. {Rodr{\'\i}guez-Garc{\'\i}a} {et~al.}(2022){Rodr{\'\i}guez-Garc{\'\i}a}, {Nieves-Chinchilla}, {G{\'o}mez-Herrero}, {Zouganelis}, {Vourlidas}, {Balmaceda}, {Dumbovi{\'c}}, {Jian}, {Mays}, {Carcaboso}, {dos Santos}, \& {Rodr{\'\i}guez-Pacheco}}]{rodriguez-garcia_2022}
{Rodr{\'\i}guez-Garc{\'\i}a}, L., {Nieves-Chinchilla}, T., {G{\'o}mez-Herrero}, R., {et~al.} 2022, \bibinfo{title}{{Evidence of a complex structure within the 2013 August 19 coronal mass ejection. Radial and longitudinal evolution in the inner heliosphere},} \aap, 662, A45, \dodoi{10.1051/0004-6361/202142966}

\bibitem[{A. Sahade(2024)Sahade}]{scc_measure3}
Sahade, A. 2024, \bibinfo{title}{scc\_measure 3,}, 1 Zenodo, \dodoi{10.5281/zenodo.13951841}

\bibitem[{A. {Sahade} {et~al.}(2023){Sahade}, {Vourlidas}, {Balmaceda}, \& {C{\'e}cere}}]{sahade_2023}
{Sahade}, A., {Vourlidas}, A., {Balmaceda}, L.~A., \& {C{\'e}cere}, M. 2023, \bibinfo{title}{{Understanding the Deflection of the ``Cartwheel CME'': Data Analysis and Modeling},} \apj, 953, 150, \dodoi{10.3847/1538-4357/ace420}

\bibitem[{A. {Sahade} {et~al.}(2025){Sahade}, {Vourlidas}, \& {Mac Cormack}}]{sahade_2024}
{Sahade}, A., {Vourlidas}, A., \& {Mac Cormack}, C. 2025, \bibinfo{title}{{Analysis of Solar Eruptions Deflecting in the Low Corona: Influence of the Magnetic Environment},} \apj, 978, 41, \dodoi{10.3847/1538-4357/ad96ba}

\bibitem[{P.~H. {Scherrer} {et~al.}(2012){Scherrer}, {Schou}, {Bush}, {Kosovichev}, {Bogart}, {Hoeksema}, {Liu}, {Duvall}, {Zhao}, {Title}, {Schrijver}, {Tarbell}, \& {Tomczyk}}]{scherrer_2012}
{Scherrer}, P.~H., {Schou}, J., {Bush}, R.~I., {et~al.} 2012, \bibinfo{title}{{The Helioseismic and Magnetic Imager (HMI) Investigation for the Solar Dynamics Observatory (SDO)},} \solphys, 275, 207, \dodoi{10.1007/s11207-011-9834-2}

\bibitem[{A. {Thernisien}(2011){Thernisien}}]{thernisien_2011}
{Thernisien}, A. 2011, \bibinfo{title}{{Implementation of the Graduated Cylindrical Shell Model for the Three-dimensional Reconstruction of Coronal Mass Ejections},} \apjs, 194, 33, \dodoi{10.1088/0067-0049/194/2/33}

\bibitem[{A. {Thernisien} {et~al.}(2009){Thernisien}, {Vourlidas}, \& {Howard}}]{thernisien_2009}
{Thernisien}, A., {Vourlidas}, A., \& {Howard}, R.~A. 2009, \bibinfo{title}{{Forward Modeling of Coronal Mass Ejections Using STEREO/SECCHI Data},} \solphys, 256, 111, \dodoi{10.1007/s11207-009-9346-5}

\bibitem[{A.~F.~R. {Thernisien} {et~al.}(2006){Thernisien}, {Howard}, \& {Vourlidas}}]{thernisien_2006}
{Thernisien}, A.~F.~R., {Howard}, R.~A., \& {Vourlidas}, A. 2006, \bibinfo{title}{{Modeling of Flux Rope Coronal Mass Ejections},} \apj, 652, 763, \dodoi{10.1086/508254}

\bibitem[{V.~S. {Titov} {et~al.}(2018){Titov}, {Downs}, {Miki{\'c}}, {T{\"o}r{\"o}k}, {Linker}, \& {Caplan}}]{titov_2018}
{Titov}, V.~S., {Downs}, C., {Miki{\'c}}, Z., {et~al.} 2018, \bibinfo{title}{{Regularized Biot-Savart Laws for Modeling Magnetic Flux Ropes},} \apjl, 852, L21, \dodoi{10.3847/2041-8213/aaa3da}

\bibitem[{T. {T{\"o}r{\"o}k} {et~al.}(2018){T{\"o}r{\"o}k}, {Downs}, {Linker}, {Lionello}, {Titov}, {Miki{\'c}}, {Riley}, {Caplan}, \& {Wijaya}}]{torok_2018}
{T{\"o}r{\"o}k}, T., {Downs}, C., {Linker}, J.~A., {et~al.} 2018, \bibinfo{title}{{Sun-to-Earth MHD Simulation of the 2000 July 14 {\textquotedblleft}Bastille Day{\textquotedblright} Eruption},} \apj, 856, 75, \dodoi{10.3847/1538-4357/aab36d}

\bibitem[{C. {Verbeke} {et~al.}(2019){Verbeke}, {Pomoell}, \& {Poedts}}]{verbeke_2019}
{Verbeke}, C., {Pomoell}, J., \& {Poedts}, S. 2019, \bibinfo{title}{{The evolution of coronal mass ejections in the inner heliosphere: Implementing the spheromak model with EUHFORIA},} \aap, 627, A111, \dodoi{10.1051/0004-6361/201834702}

\bibitem[{A. {Vourlidas}(2014){Vourlidas}}]{vourlidas_2014}
{Vourlidas}, A. 2014, \bibinfo{title}{{The flux rope nature of coronal mass ejections},} Plasma Physics and Controlled Fusion, 56, 064001, \dodoi{10.1088/0741-3335/56/6/064001}

\bibitem[{J.-P. {Wuelser} {et~al.}(2004){Wuelser}, {Lemen}, {Tarbell}, {Wolfson}, {Cannon}, {Carpenter}, {Duncan}, {Gradwohl}, {Meyer}, {Moore}, {Navarro}, {Pearson}, {Rossi}, {Springer}, {Howard}, {Moses}, {Newmark}, {Delaboudiniere}, {Artzner}, {Auchere}, {Bougnet}, {Bouyries}, {Bridou}, {Clotaire}, {Colas}, {Delmotte}, {Jerome}, {Lamare}, {Mercier}, {Mullot}, {Ravet}, {Song}, {Bothmer}, \& {Deutsch}}]{wuelser_2004}
{Wuelser}, J.-P., {Lemen}, J.~R., {Tarbell}, T.~D., {et~al.} 2004, in Society of Photo-Optical Instrumentation Engineers (SPIE) Conference Series, Vol. 5171, Telescopes and Instrumentation for Solar Astrophysics, ed. S.~{Fineschi} \& M.~A. {Gummin}, 111--122, \dodoi{10.1117/12.506877}

\end{thebibliography}
\bibliographystyle{aasjournalv7}



\end{document}